\documentclass[conference]{IEEEtran}
\IEEEoverridecommandlockouts
\usepackage{graphicx}
\usepackage[caption=false]{subfig}
\usepackage[hidelinks]{hyperref}
\hypersetup{
    colorlinks=true,
    linkcolor=black,
    filecolor=ForestGreen,      
    urlcolor=blue,
    citecolor = black
}

\begin{document}
%
% paper title
% Titles are generally capitalized except for words such as a, an, and, as,
% at, but, by, for, in, nor, of, on, or, the, to and up, which are usually
% not capitalized unless they are the first or last word of the title.
% Linebreaks \\ can be used within to get better formatting as desired.
% Do not put math or special symbols in the title.
\title{Framework for Automatic PCB Marking Detection and Recognition for Hardware Assurance
}
% author names and affiliations
% use a multiple column layout for up to three different
% affiliations
\author{
\IEEEauthorblockN{
Olivia P. Dizon-Paradis,
Daniel E. Capecci, 
Nathan T. Jessurun,\\
Damon L. Woodard,
Mark M. Tehranipoor,
and Navid Asadizanjani}

\IEEEauthorblockA{
Electrical and Computer Engineering Department, University of Florida\\
Gainesville, FL, USA\\
Email: paradiso@ufl.edu\\}

\thanks{DISTRIBUTION STATEMENT A.  Approved for public release: distribution is unlimited.}
}

\maketitle

\begin{abstract}
A Bill of Materials (BoM) is a list of all components on a printed circuit board (PCB). Since BoMs are useful for hardware assurance, automatic BoM extraction (AutoBoM) is of great interest to the government and electronics industry. To achieve a high-accuracy AutoBoM process, domain knowledge of PCB text and logos must be utilized. In this study, we discuss the challenges associated with automatic PCB marking extraction and propose 1) a plan for collecting salient PCB marking data, and 2) a framework for incorporating this data for automatic PCB assurance. Given the proposed dataset plan and framework, subsequent future work, implications, and open research possibilities are detailed. 
\\

\end{abstract}
\renewcommand\IEEEkeywordsname{Keywords}

\section{Introduction}
A Bill of Materials (BoM) is a complete inventory of all components present on a printed circuit board (PCB), e.g. resistors, capacitors, and integrated circuits (ICs). BoMs can be used in a variety of applications in the domain of PCB reverse engineering (e.g. analysis of foreign, competitor, and legacy devices), hardware assurance (e.g. vulnerability identification and hardware Trojan detection), industrial assessment (e.g. manufacturing defect detection, revenue estimation), and academia (e.g. technology trend research)~\cite{quadir_survey_2016}. 

Consequently, frameworks to automatically extract a PCB's BoM using optical images were introduced in~\cite{azhagan_2019} and~\cite{paradis_2020}. In this study, the role of PCB markings, i.e. logos and text, is detailed in the context of these frameworks. 

\begin{figure}[!htbp]
    \centering
    \includegraphics[width=0.75\linewidth]{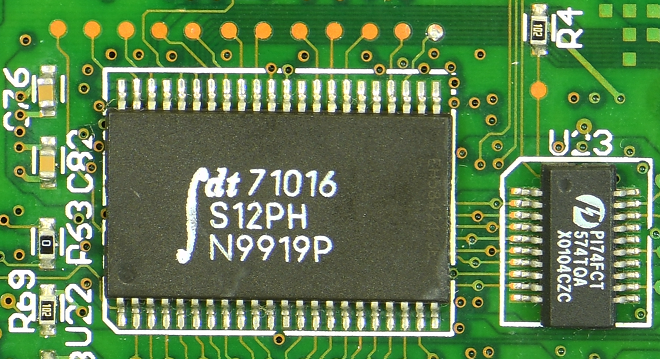}
    \caption{Example of PCB markings, i.e. text and logos.}
    \label{fig:PCB}
\end{figure}

PCBs are used in critical government, military, and biomedical infrastructures and are often mass-manufactured with hundreds or thousands of components. Therefore, 100\% accuracy is highly desirable, as even a small error in AutoBoM accuracy means the potential misrepresentation of hundreds of PCB components. Subject matter experts (SMEs) use PCB markings, such as those shown in Fig.~\ref{fig:PCB}, as a vital indicator for classifying and identifying each component to construct a BoM. 

In this work, we propose PCB markings as a route to achieve higher AutoBoM accuracy. For example, components may be associated with reference designators, or on-board text which typically consist of one or two letters followed by a number, e.g. R69, C76, and U22 in Fig.~\ref{fig:PCB}. These reference designators can be used to help classify each component. As represented in the figure, components with reference designators beginning with an "R" are typically resistors, "C" are capacitors, "U" are ICs, etc. In addition to on-board markings, on-component text and manufacturer logos, such as those on the ICs in Fig.~\ref{fig:PCB}, provide critical information to specifically identify the exact component. To leverage on-board and on-component information for component identification, SMEs must be familiar with reference designation standards (e.g. ASME Y14.44-2008~\cite{asme_2008} and IEEE 315-1975~\cite{ieee_1975}), as well as manufacturer logos, computer-aided design (CAD) design rule checks, and best practices. Such PCB design rules may be used for automatic BoM extraction by informing parameter constraints and priors.  

Several algorithms exist for automatic detection and recognition of text and logos in a general sense. At present, few existing algorithms are specifically meant to address PCB markings~\cite{Li_Neullens_Breier_Bosling_Pretz_Merhof_2014, Iano_Bonello_Neto_2020}. Parameter-tuning, method fusion, and transfer learning methodologies could improve accuracy for detection and recognition of text, but they are not as effective for detection and recognition of logos because many types of logos are unique to the electronics domain. Moreover, there is a lack of annotated PCB marking data in the public domain to train and/or evaluate potential new algorithms~\cite{Pramerdorfer_Kampel_2015, lu_2020}. The FICS-PCB dataset contains annotations for components, on-board text, and on-component text, but the text is annotated in a format inconsistent with traditional optical character recognition (OCR) techniques and logos are not identified~\cite{lu_2020}. To address these research gaps, we propose 1) a plan for PCB marking data collection and 2) a framework for incorporating marking data for automatic PCB assurance. 

The rest of this paper is organized as follows. First, Sec.~\ref{sec:challenges} elaborates on the real-word challenges to consider for automatic PCB marking detection and recognition. Then, Sec.~\ref{sec:data} presents a proposed plan for data collection of PCB marking data. Sec.~\ref{sec:proposed framework} details a framework for incorporating this data for automated PCB assurance. Finally, Sec.~\ref{sec:conclusion} concludes the study with a summary of the paper's key takeaways and future research directions.

\section{Challenges}\label{sec:challenges} 
As in other domains, there are several difficulties for automatic PCB marking detection and recognition~\cite{ye_2014}. Examples of difficult cases are shown in Fig.~\ref{fig:challenges}. 

\subsection{High Intra-class Variation}
There are many different form factors each character or logo may exhibit. In other words, instances of each marking may display a wide variety of colors, shapes, and textures. For example, the character ``4" on the IC in Fig.~\ref{subfig:example a}, the capacitor in \ref{subfig:example b}, and the resistor in \ref{subfig:example d} all appear different. Though methods to detect text in various fonts and orientations is improving \cite{Coates2011-va, Shi2017-vw}, many of these algorithms are tuned to images with only a few instances of markings per image. This is not the case for PCBs, which may consist of hundreds or thousands of densely-packed markings. In addition, there are also other characteristics of PCB images that need to be tested. 

\subsection{Low Inter-class Variation}
\begin{figure}%[!htbp]
  \centering
  \subfloat[\label{subfig:example a}]{%
    \includegraphics[width=0.31\linewidth, height=0.31\linewidth]{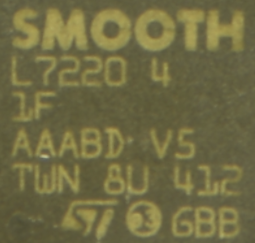}%
  }
  \hfill
  \subfloat[\label{subfig:example b}]{%
    \includegraphics[width=0.31\linewidth, height=0.31\linewidth]{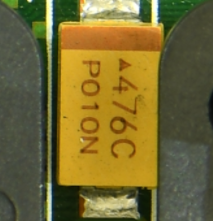}%
  }
  \hfill
  \subfloat[\label{subfig:example c}]{%
    \includegraphics[width=0.31\linewidth, height=0.31\linewidth]{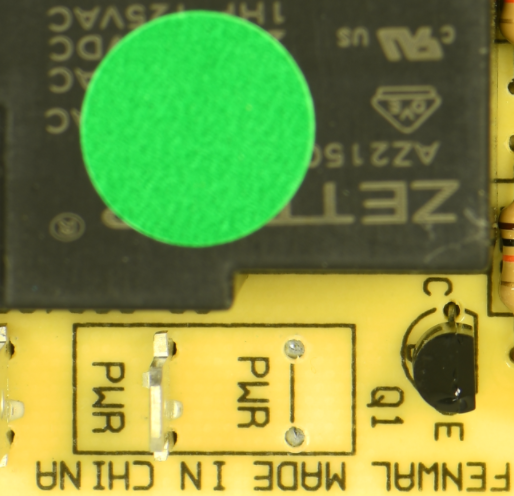}%
  }
  \hfill
  \subfloat[\label{subfig:example d}]{%
    \includegraphics[width=0.31\linewidth, height=0.31\linewidth]{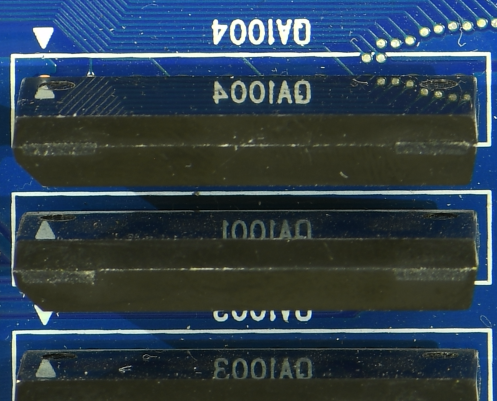}%
  }
  \hfill
  \subfloat[\label{subfig:example e}]{%
    \includegraphics[width=0.31\linewidth, height=0.31\linewidth]{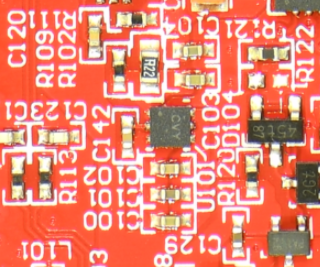}%
  }
  \hfill
  \subfloat[\label{subfig:example f}]{%
    \includegraphics[width=0.31\linewidth, height=0.31\linewidth]{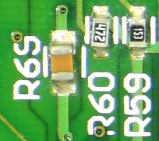}%
  }
  \caption{Examples of challenging cases for Automatic PCB Marking Detection and Recognition.}
  \label{fig:challenges}
\end{figure}

As each marking may exhibit different form factors, there are also instances where one marking may be mistaken for another due to low inter-class variation. For instance, the STMicroelectronics logo in Fig.~\ref{subfig:example a}, letter ``O" in \ref{subfig:example b}, and ``IN" in \ref{subfig:example c} may be mistaken for the characters ``ST", number ``0", and ``NI", respectively. Note that, though the surrounding text in Fig.~\ref{subfig:example c} provides insight into the upside-down orientation of ``IN", there are cases where such context is not available and the ambiguous orientation can cause misclassification errors. 

\subsection{Imbalanced Classes}
Another factor to consider is the uneven number of instances for each class of PCB marking.  For example, since PCBs generally possess more resistors and capacitors than ICs, the associated reference designators ``R" and ``C" are more common than ``U" (Fig.~\ref{subfig:example e}). Furthermore, logos pose additional challenges because there are often significantly more characters than logos on PCBs. Even when considering only logos, there may still be class imbalances between logos on common off-the-shelf components versus custom-made components. Class imbalances, if unaddressed, can heavily bias recognition algorithms and performance metrics. 

\subsection{Imaging Challenges}
Complications may also arise due to the way the PCBs are imaged. Noise due to camera sensors, lighting, or the internal digital camera image processing pipeline can introduce artifacts and/or obfuscate the markings. Markings may not be properly rendered if the camera resolution is too poor, while the image data may be difficult to efficiently process if the resolution is too high. Uneven lighting conditions can introduce shine, shadows, and other color variations, which thereby increase the intra-class variation of the different markings. For instance, in Fig.~\ref{subfig:example c}, the IC text is relatively out-of-focus as the camera's sensors are focused on the board text. For each imaging parameter, it is important to consider the possible trade-offs. For example, shine may obscure some markings, but it can also be beneficial for texture analysis-based detection of marking defects and component localization~\cite{Ibrahim2010-vs}. Different imaging setups have their own benefits and limitations in terms of performance and computation time/speed.

\subsection{Physical Challenges}
Challenges may also arise due to physical properties of the PCB components and the board itself. Text and logos can fade due to age, wear-and-tear, and/or damage (e.g. Fig.~\ref{subfig:example a}). Markings can also be occluded by components, remnants of PCB packaging, and/or stickers as in Fig.~\ref{subfig:example c}. Fig.~\ref{subfig:example d} shows an example of false markings due to reflection of a true marking off a smooth component. Markings can also be broken or interrupted by components, other markings, traces, and vias as in Fig.~\ref{subfig:example f}. Physical challenges such as these can often create false positive detections, false negatives, or misclassifications. Though some physical challenges can be addressed by altering the image acquisition setup, some are difficult to overcome without physically altering the board.

\subsection{Complex Component-Marking Associations}
After markings have been detected and recognized, the next step for hardware assurance would be to relate the markings to the different PCB components. Though it is relatively straightforward to relate board markings to the devices they are printed on, it is nontrivial to relate board markings to their respective components. In an ideal situation, each component is associated with exactly one board marking, which is a reference designator, and each component-designator pair is located closest to each other (i.e. they are unambiguously related). However, in practice, there are components associated with multiple board markings or none at all, and vice versa. For example, a reference designator such as ``R1-3" may correspond to three components, with R2 and R3 implicitly defined. Moreover, due to the density of the components and board markings, it is not always the case that the closest component-designator pairs are associated. For example, the light brown capacitor in the upper left of Fig.~\ref{subfig:example e} is not associated with ``R102" or ``R111", but rather ``C120". Such complex associations require the use of board context and domain knowledge to infer the true relationships. 

\subsection{Evolving Scope}
In addition, not only is the problem of automatic PCB marking detection and recognition challenging, but the problem is constantly evolving. As technology advances, markings tend to become smaller and more compact along with similar trends of smaller components and more compact PCBs. Along with technological changes, manufacturing contexts are also constantly shifting. New manufacturers, mergers and acquisitions, and development of new logo variations over time (e.g. to accommodate smaller components or different layouts) can all affect the types of logos that appear on a PCB. In other words, the number of PCB marking classes is changing over time and the necessary database to properly map these symbols grows. In addition, other variabilities such as hardware Trojans, design errors, and manufacturing defects must be accounted for. The ``R69" resistor-like reference designator for the capacitor in Fig.~\ref{subfig:example f} is one such example. Since the problem is constantly evolving, there is a need from the hardware assurance community for increased collaboration and a continually updated dataset, as proposed in the following section. 

\section{Proposed Data Collection Plan}\label{sec:data} 
As detailed in Sec.~\ref{sec:challenges}, there are several challenges for automatic PCB  marking detection and recognition, many of which can be addressed by incorporating more data and prior knowledge. However, the lack of available PCB marking data in the hardware assurance community makes this solution currently infeasible. Furthermore, collecting this data is expensive and time-consuming. Hence, such PCB marking datasets tend to be quite limited in either the number of distinct classes or instances of each class. Though there are many datasets of text and logos in other domains (e.g. text in natural scenes~\cite{zhang2013text}, logos of commercial brands~\cite{wang2020logodet} and clothing~\cite{liu2021cbl}, etc.), few are representative of all the challenges present in the PCB domain. Therefore, a plan for collecting PCB marking data is proposed. 

To ensure PCB assurance algorithms are robust, there should be a method of color and scale normalization incorporated in the workflow. For example, a color checker consisting of an array of different color swatches with known RGB values and a ruler can be used to normalize the lighting conditions and resolution of images taken with different cameras. Such normalization of data taken with different sensors facilitates the development of reproducible algorithms. This, in turn, helps address the high intra-/low inter-class variations and the imaging challenges. 

Consistent with existing text and logo datasets from other domains, PCB markings should be annotated by word~\cite{ye_2014}. Though semantic marking annotations may improve algorithm performance, there is a considerable time/performance trade-off. Many stock marking recognition algorithms first find words, then localize characters, and finally classify the characters~\cite{zhang2013text}. Though accuracy of these stock algorithms may not be very high, they can be used to help mitigate the time/performance trade-off of semantic annotation. 

Since there is a wide and ever-increasing variety of logos, a lookup table of logos should be used to ensure classes are annotated as consistently as possible. Consistent class labelling can address the challenges from arbitrary additions of unnecessary classes such as low inter-class variation between equivalent logos and potential class imbalances from over-representing the same logo under different class names. 

Other information that should be annotated include whether the marking is defective, its orientation, and any devices associated with it. Noting the defects, especially the type (e.g. wearing versus misprints), helps algorithms better model the effects of such defects on inter-class variation. Such algorithms are useful for defect detection, a critical aspect of PCB assurance. Orientation tracking is used to train and test algorithms to resolve ambiguous cases (e.g. resistor text ``221" can look like ``122" upside down, and vice versa). These annotations can be verified (e.g. against a published list of standard resistor values) to ensure an accurate grammar model is produced. These algorithms are useful for specifically identifying the different PCB components. Finally, tracking associations between markings and devices is useful for training and testing algorithms to determine implicit, ambiguous, or other complex relations. Such algorithms assist in classifying and identifying the different PCB components. 

\begin{figure*}[!htbp]
    \centering
    \includegraphics[width=.75\linewidth]{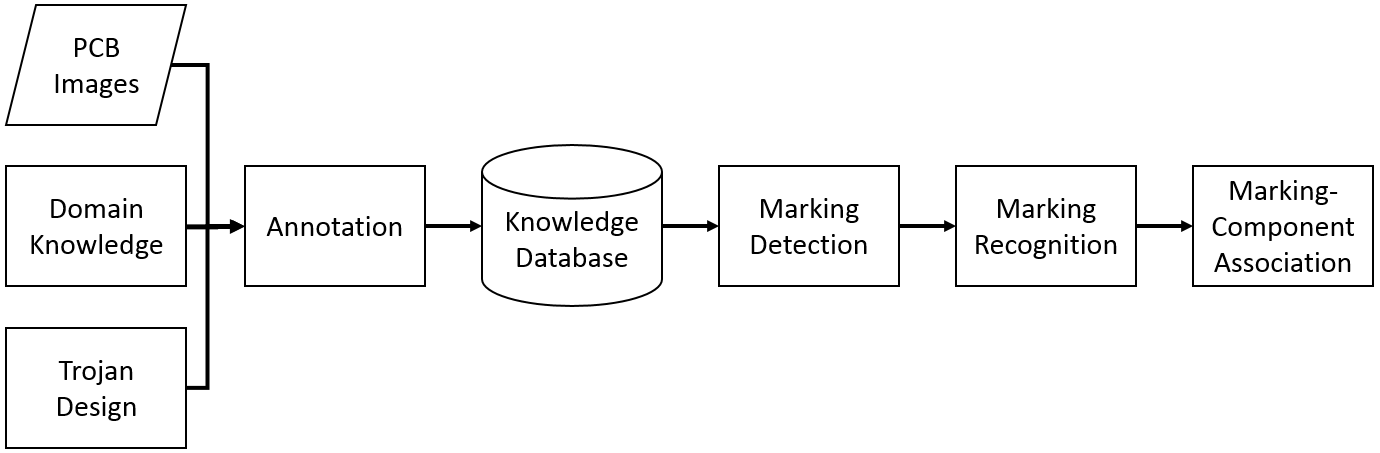}
    \caption{Proposed framework for automatic PCB marking detection and recognition for hardware assurance.}
    \label{fig:proposedFrmwrk}
\end{figure*}
    
Note that, though additional information such as font size, text color, and board color could also improve PCB marking detection and recognition algorithms, they are not specifically mentioned in the data collection plan. These properties are excluded because many can be automatically extracted given the information obtained from the above data collection plan combined with simple image processing and/or computer vision algorithms. Moreover, the proposed data collection plan is intended as a starting point to facilitate research and collaboration within the hardware assurance community. Hence, additional details for improved general recognition will be incorporated as the problems are more deeply understood. After the marking data has been collected, it will be used for PCB hardware assurance. Toward this end, we propose the following PCB marking detection and recognition framework in the next section. 

\section{Proposed Framework}\label{sec:proposed framework}

The proposed framework for automatic PCB marking detection and recognition for hardware assurance is shown in Fig.~\ref{fig:proposedFrmwrk}. Individual processes are detailed in the following subsections. 

\subsection{Domain Knowledge}
Subject matter expertise concerning PCB markings from design rules, company standards, physical constraints, etc. can provide helpful details about the nature of PCB components. Examples of such details include standard resistor values, lists of component manufacturer logos, and rules for reference designator placement. \textit{Domain Knowledge} consists of collecting this information so that more accurate annotations (and subsequently, automated PCB analysis algorithms) can be produced. For example, insights for reference designator placement can help algorithms classify and identify the associated components. In addition, incorporation of domain knowledge as a prior to PCB assurance algorithms may improve performance while requiring less data. 

\subsection{Trojan Design}
An important factor in reliably detecting Trojans is understanding their various appearances. Hardware Trojans, which are designed to go undetected, may possess false markings to disguise themselves. By noting these marking peculiarities (which may appear different from those used by the true manufacturers) and researching state-of-the-art Trojan designs, the PCB assurance framework can maintain a regularly updated library of adversarial attacks, as in~\cite{noauthor_trust-huborg_nodate}. As a result, hardware assurance algorithms can cross-reference markings and their associated components against this library to better identify malicious alterations. 

\subsection{Annotation}
During \textit{Annotation}, SMEs leverage domain knowledge and Trojan design principles to label PCB images. Here, text and component logos are labeled along with their associated PCB components to produce ground truth information. The ground truth is then used to train machine learning algorithms. As discussed in Sec.~\ref{sec:data}, a plan for collecting PCB marking data is proposed which details information to annotate. When the domain of automatic PCB assurance has matured, SME involvement may be reduced from fully-manual efforts to semi-supervised efforts. This transition to semi-supervised efforts may help address the challenges in Sec.~\ref{sec:challenges} by dramatically increasing the amount of data available for PCB assurance research.

\subsection{Knowledge Database}
All image, marking annotations, component annotations, and associated metadata (e.g. Trojan exploits and domain knowledge) are then stored in the \textit{Knowledge Database}. This data will be readily accessible as both reference material and ground truth for algorithm parameter tuning and performance evaluation. Information stored in this database will encompass a wide variety of markings from different component types, manufacturers, generations, and more. A comprehensive range of represented markings and associated devices ensures PCB assurance algorithms are not over-tuned to a select few PCBs. In addition, this \textit{Knowledge Database} of information accumulated over time can help address the challenge of an evolving scope of PCB assurance. 

\subsection{Marking Detection}
Information from the previous stages are then processed during \textit{Marking Detection} and leveraged for algorithm training, cross-referencing, and result evaluation. By the end of this procedure, a list of regions where markings may exist in the PCB images is output.

\subsection{Marking Recognition}
Each region of interest (ROI) of potential markings obtained from \textit{Marking Detection} is then recognized. This involves determining the specific characters or logos within each ROI. Text can be recognized by the characters they consist of, while logos can be recognized by their respective manufacturer. 

\subsection{Marking-Component Association}
Finally, recognized markings obtained from \textit{Marking Recognition} are then associated with PCB components such as resistors, capacitors, ICs, inductor, no component, etc. Both on-board and on-device markings provide valuable insight to classify and identify the associated components.

\subsubsection{On-board Markings}
Markings such as on-board reference designators provide valuable insight to classify the associated components. For example, components with reference designators beginning with an "R" are typically resistors, "C" are capacitors, "U" are ICs, etc. This can be helpful in ambiguous cases where a component's appearance straddles the decision boundary between two classes (e.g. a dark gray inductor may look similar to a resistor). However, it is important to note that the process of classifying components using reference designators should be iterative to account for manufacturing defects and hardware Trojans. For example, the capacitor in Fig.~\ref{subfig:example f} has the resistor-like reference designator ``R69". Here, a marking recognition algorithm may be certain the component is a resistor because of the ``R69" reference designator, but a component recognition algorithm may be certain the component is a capacitor based on its appearance. In this case, the confidences of both recognition algorithms should be considered and unclear cases should be flagged for review by an SME. In this way, markings are not intended to be the only source of information used to classify the components, only a way to verify and refine predictions with higher confidence. 

\subsubsection{On-device Markings}
Markings such as on-device text and logos provide insight for uniquely identifying specifics about the associated components, such as a parameter values, serial numbers, and logos. For instance, a component classified as a ``resistor" with a ``221" on it may be identified as a ``220 Ohm resistor". Note that domain knowledge was used here to infer the last digit is an exponent, i.e. ``221" resistor text = $22\times10^1 = 220$ Ohms. Similarly, an object classified as an ``IC" may be identified by a logo and serial number, which can be used to search for the respective specifications sheet published by the manufacturer. As with classifying components using markings, the process of identifying them should also be iterative to account for manufacturing defects and hardware Trojans. In this way, markings are also not intended to be the only source of information used to identify the components, only a way to verify and refine predictions with higher confidence.

\section{Conclusion}\label{sec:conclusion}
This paper suggests text and logo information can be useful for automated PCB assurance. However, at present, extracting and using this marking information is difficult with the limited amount of data that is publicly available. This paper elaborates on the real-word challenges with PCB marking data and details why it is important to develop a dataset specially for PCB text and logos. We therefore propose a plan for collecting PCB marking data and a framework for incorporating this marking data for automatic PCB assurance. The proposed data collection plan and framework are intended as a starting point to facilitate research and collaboration within the hardware assurance community. 

Future work includes collecting and annotating PCB data, as outlined in Sec.~\ref{sec:data}, and incorporating this data for automated PCB assurance, as detailed in Sec.~\ref{sec:proposed framework}. From there, the results, benefits, and limitations of algorithms incorporating and not incorporating marking information will be compared. These resources and tools will enable progress toward accurate, automated hardware assurance. 

\section*{References}

\def\refname{}

% \bibliographystyle{IEEEtran}
% \bibliography{references.bib}

\end{document}